\documentclass[pre,showpacs,showkeys,floatfix,nofootinbib,reprint]{revtex4-1}

\usepackage{graphicx}
\usepackage{amsmath}
\usepackage{amsfonts}
\usepackage{bm}
\usepackage{bbm}
\usepackage{url}
\usepackage{amssymb}
\usepackage{mathrsfs}
\usepackage{dsfont}
\usepackage{subfigure}
\usepackage{paralist}

\usepackage{stmaryrd}

\newcommand{\zero}{\bm{0}}
\newcommand{\vel}{\bm{v}}
\newcommand{\acc}{\bm{a}}
\newcommand{\n}{\bm{n}}
\newcommand{\ut}{\bm{t}}
\newcommand{\tp}{\bm{t}^+}
\newcommand{\tm}{\bm{t}^-}

\newcommand{\ex}{\bm{e}_x}
\newcommand{\ey}{\bm{e}_y}
\newcommand{\jump}[1]{\mbox{$\llbracket #1\rrbracket$}}
\newcommand{\tzero}{\vartheta_0}
\newcommand{\tone}{\vartheta_1}
\newcommand{\vt}{\vartheta}
\newcommand{\lv}{\lambda\vel}
\newcommand{\lvtwo}{\lambda v^2}
\newcommand{\lvthree}{\lambda v^3}
\newcommand{\Ws}{W_\mathrm{s}}
\newcommand{\Wa}{W_\mathrm{a}}
\newcommand{\taut}{\tau\ut}
\newcommand{\sspeed}{\dot{s}_0}
\newcommand{\force}{\bm{f}}
\newcommand{\degree}{^\circ}
\newcommand{\Neptune}{\Psi}
\newcommand{\Force}{\bm{\Phi}}
\newcommand{\pone}{p_1^+}
\newcommand{\pzero}{p_0^-}
\newcommand{\taup}{\tau^+_1}
\newcommand{\taum}{\tau^-_0}

\begin{document}
\title{Dissipative shocks in a chain fountain}\thanks{This paper is dedicated to the memory of Piero Villaggio, who died on the 4th of January 2014, aged 81.}
\author{Epifanio G. \surname{Virga}}
\email[e-mail: ]{eg.virga@unipv.it}
\affiliation{ Dipartimento di Matematica, Universit\`a di Pavia, Via Ferrata 5, I-27100 Pavia, Italy}

\date{\today}

\begin{abstract}
The fascinating and anomalous behaviour of a chain that instead of falling straight down under gravity, first rises and then falls, acquiring a steady shape in space that resembles a fountain's spray, has recently attracted both popular and academic interest. The paper presents a complete mathematical solution of this problem, whose distinctive feature is the introduction of a number of dissipative shocks which can be resolved exactly.
\end{abstract}



\maketitle
\section{Introduction}\label{sec:intro}
It is amazing how a topic as classic as string mechanics, which has been tended for centuries, may still surprise and challenge us with unsolved problems. This is the case for the \emph{chain fountain}, the anomalous behaviour of a chain falling under gravity which has already fascinated the millions who viewed the movie of S. Mould~\cite{mould:self}, myself included.\footnote{A similar phenomenon had indeed been documented earlier by J.~A. Hanna and H.~King~\cite{hanna:instability}.} In words, one end of a long metal chain sitting in a pot is raised above the pot's rim and let fall towards the floor. Contrary to everybody's expectation, in most experimental circumstances, the chain instead of falling straight down climbs appreciably higher than the pot's rim before falling, thus drawing a steady curve in space which resembles a fountain's spray.

Chains, cords, and strings\footnote{These three nouns will be regarded here as synonyms, as our mathematical development will be equally applicable to all these model bodies.} have been the object of theoretical mechanics since 1614--1615 when, according to Truesdell~\cite[p.\,24]{truesdell:rational}, Beeckman is likely to have found that the equilibrium shape of a string uniformly loaded along a line (such as a suspension bridge) is a parabolic arc.\footnote{Erroneously, such a conclusion was also reached by Galilei in 1638~\cite[pp.\,369--370]{galilei:discorsi} for the equilibrium shape of a homogeneously heavy cord. The correct catenary solution was found subsequently by Leibniz, Huygens, and James Bernoulli, apparently independently. The reader may also consult \cite[p.\,303]{villaggio:mathematical} for a concise historical account.} James Bernoulli is usually credited with the discovery of the analytic form for the equilibrium shape of a chain under gravity, called the \emph{catenary} from the Latin word for chain.

More recently, that is to say, in the late 19th century, it was remarked by Airy~\cite{airy:mechanical} that the catenary is also the steady shape of a string being drawn at constant speed, the passage from equilibrium to steady motion only resulting in offsetting uniformly the string's tension by a quantity proportional to the velocity square.\footnote{It is remarkable how such a discovery was indeed prompted by the failure of the first attempt at depositing a transatlantic telegraphic cable, an engineering problem which was still an object of study a century later \cite{zajac:dynamics}.} Plenty of historical remarks illuminating the pedigree of this problem can also be found in the papers by Biggins and Warner~\cite{biggins:understanding,biggins:growth}, upon which I further elaborate in this work.

That an arc of inverted catenary is also the steady shape of a falling chain there can be no doubt. The boundary conditions to which such an arc is subjected are essential to the understanding of the dynamics of chain fountains, and here opinions may differ. Biggins and Warner~\cite{biggins:understanding,biggins:growth} proposed that the chain is actually lifted up at the detach point and pulled down at the deposition point, the corresponding forces being produced by the pot and the floor, respectively. Several arguments are given in \cite{biggins:understanding,biggins:growth} that reduce both forces to elementary physical mechanisms involving the nature and shape of the links constituting the chain. Suggestive as these arguments may be, they eventually result in separate constitutive assumptions for the tension $\tau$ of the chain at the pickup and putdown points, which read as
\begin{equation}\label{eq:BW_tensions}
\tau_0=(1-\alpha)\lambda v^2,\qquad\tau_1=\beta\lambda v^2,
\end{equation}
respectively, where $\lambda$ is the mass per unit length of the chain, $v$ is the velocity at which it is drawn, and $0\leqq\alpha\leqq1$ and $\beta\geqq0$ are dimensionless parameters.\footnote{For dimensional reasons, both tensions must be proportional to $\lambda v^2$. In \eqref{eq:BW_tensions}, where both (12) and (13) of \cite{biggins:growth} are combined, the dimensionless parameters $\alpha$ and $\beta$ appear to be constitutive of both the chain and the environment it comes in contact with.} According to Biggins and Warner's explanation, the negative contribution to $\tau_0$, $-\alpha\lambda v^2$,  amounts precisely to the reactive upward force exerted by the pot onto the chain's link being set in motion. Contrariwise, the tension $\tau_1$, which is exerted on the chain by the heap of links being freely collected at the foot of the fountain, is not required to vanish, as if the terminal link were still freely flying before the impact with the floor, in accord with some recent studies \cite{hamm:weight,grewal:chain}, which have already advanced theoretically and also confirmed experimentally such a hypothesis.

Here I do not question the validity of \eqref{eq:BW_tensions}, but I want to derive \eqref{eq:BW_tensions} from first principles, expressing both $\alpha$ and $\beta$ in terms of a single internal constitutive parameter pertaining only to how the chain is made. The way to achieve this will be by regarding both the pickup and putdown points as standing shocks that dissipate energy at the rate dictated by a classical law for internal impacts. To this end, we need first recall the theory of shocks in one-dimensional continua. Luckily this task is made easy by a paper of O'Reilly and Varadi~\cite{reilly:treatment} who, elaborating on earlier work of Green and Naghdi~\cite{green:thermodynamics,green:note,green:derivation,green:thermal}, proposed an elegant and rather comprehensive theory, which in Sec.~\ref{sec:shocks} is applied to the case at hand. Section~\ref{sec:shocks}, which is the heart of the paper, is split in several subsections to make it easier for the reader retrace the different elements of the theory developed here. The discussion proposed in the final Sec.~\ref{sec:discussion} draws a closer comparison with the theory of Biggins and Warner and indicates further possible applications of the present theory.

\section{Dissipative Shocks}\label{sec:shocks}
Think of a chain as an \emph{inextensible} string with uniform mass density $\lambda$ per unit length, parameterized in the reference configuration by the arc-length $s$. The position in space occupied by a material point of the string is represented  by the mapping $p=p(s,t)$. Here $s$, which designates the convected variable, could as well be used to designate the arc-length in the present configuration. Correspondingly, the velocity $\vel$ is defined by $\vel:=\dot{p}$, where a superimposed dot represents differentiation with respect to time $t$. Similarly, $\acc:=\dot{\vel}$ is the acceleration. Let $\force$ denote the \emph{external} force acting per unit length of the string and $\tau\geqq0$ the internal tension that arises as a reaction to the inextensibility constraint. The balance of linear momentum along any smooth arc of the string is expressed by
\begin{equation}\label{eq:smooth_momentum_balance}
\lambda\acc=\force+(\taut)^\prime,
\end{equation}
where a prime $^\prime$ denotes differentiation with respect to $s$ (see, for example, \cite[Sec.\,34]{villaggio:mathematical}).

\subsection{Shock Equations}\label{sec:shock_equations}
\begin{figure}[ht]
  \includegraphics[width=.6\linewidth]{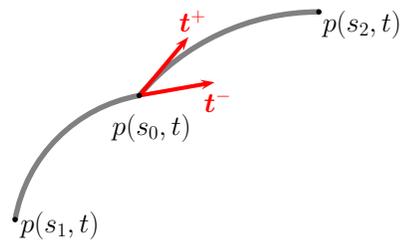}
  \caption{
  The present shape at time $t$ of the string. The points $p(s_1,t)$ and $p(s_2,t)$, with $s_2>s_1$, delimit the arc under consideration. The point $p(s_0,t)$ is a singular point, where the unit tangent $\ut$ is discontinuous, with traces $\tp$ and $\tm$ on the two sides.}
\label{fig:shock}
\end{figure}

In this context, a \emph{shock} propagating along the string is described by a function, $s_0=s_0(t)$, identifying the point in the reference configuration carrying a discontinuity in speed (and acceleration). Specifically, we assume that $p$ is continuous at $s_0$, because the string breaks nowhere, but $\vel$ is discontinuous. Similarly, both the unit tangent $\ut$, the principal unit normal $\n$, and the curvature $c$ of the curve representing the present shape (at time $t$) of the string are discontinuous at $p(s_0(t),t)$, as illustrated in Fig.~\ref{fig:shock}.
We shall call $p(s_0,t)$ a \emph{singular point}. We denote by $\tp$ and $\tm$ the two limiting values of $\ut$ across a singular point. Here and below, for any quantity $\Neptune$, superscripts $^\pm$ refer to the traces of $\Neptune$ across $p(s_0,t)$ from the sides of increasing and decreasing $s$, respectively. Also, we shall employ the customary notation $\jump{\Neptune}:=\Neptune^+-\Neptune^-$, for the \emph{jump} of $\Neptune$ across a singular point.

The shock speed is $\sspeed$ relative to both the reference and present shapes (as a consequence of the string's inextensibility). A kinematic compatibility condition arises for the jumps of both $\vel$ and $\acc$, as a result of the requirement that both the velocity and acceleration of the geometric point which instantaneously coincides in space with a singular point can be expressed in two different, but equivalent ways (see, for example, \cite{reilly:treatment,reilly:energetics}):
\begin{subequations}\label{eq:kinematic_compatibility}
\begin{equation}\label{eq:kinematic_compatibility_velocity}
\jump{\vel}+\sspeed\jump{\ut}=\zero,
\end{equation}
\begin{equation}\label{eq:kinematic_compatibility_acceleration}
\jump{\acc}+2\sspeed\jump{\vel'}+\sspeed^2\jump{c\n}+\ddot{s}_0\jump{\ut}=\zero.
\end{equation}
\end{subequations}

The balance of linear momentum for an arbitrary small arc enclosing a singular point (that is, for $s_2\to s_0^+$ and $s_1\to s_0^-$ in Fig.~\ref{fig:shock}), requires that
\begin{equation}\label{eq:jump_conditions_linear_momentum}
\jump{\taut}+\sspeed\jump{\lv}+\Force=\zero,
\end{equation}
where $\Force$ is the concentrated supply of momentum that must be provided at a singular point to sustain  the shock. In a similar way (see again \cite{reilly:treatment,reilly:energetics} for more details), the energy balance at a singular point results into the following equation,
\begin{equation}\label{eq:jump_conditions_energy}
\jump{\taut\cdot\vel}+\frac12\lambda\sspeed\jump{v^2}+\Ws=0,
\end{equation}
where $\Ws$ is the concentrated power supply involved in the shock.\footnote{Equation \eqref{eq:jump_conditions_energy} is a specialization to the athermal case treated here of equation (2.7)$_4$ of \cite{reilly:treatment}; what here is denoted $\Ws$ was there denoted $\Phi_E$. For $\Ws<0$, energy is lost in the shock. According to Sommerfeld's book \cite{sommerfeld:mechanics} (see, pp.\,28--29 and Problem I.7, pp.\,241, 257), the energy loss in chain dynamics is a concept first introduced by Lazare Carnot, the father of Sadi (this latter known for his contributions to the theory of heat), who was a writer on mathematics and mechanics (besides later becoming one of the most loyal of Napoleon's generals). See also \cite[p.\,52]{muller:history}, \cite{steiner:equations} and \cite{wong:falling}.} For a dissipative shock, $\Ws$ is negative and measures the energy lost per unit time by the internal frictions that hamper the shock as it  goes by. While in our setting the force $\Force$ will be provided through the contact of the chain with the external world, $\Ws$ is of a constitutive nature, which needs to be further specified (see Sec.~\ref{sec:shock_dissipation}). Equations \eqref{eq:jump_conditions_linear_momentum} and \eqref{eq:jump_conditions_energy} express only the mechanical balances at a singular point. The former is also known as the \emph{Rankine-Hugoniot} jump condition for one-dimensional continua \cite[p.\,29]{antman:nonlinear}.\footnote{The reader is further referred to \cite{reilly:treatment,reilly:energetics} for a general thermodynamic theory of strings, which also features an additional jump condition for the entropy imbalance. A formulation of shock waves for general three-dimensional continua can also be found in Secs.~32 and 33 of \cite{gurtin:mechanics}.}

Interesting versions of the jump conditions in \eqref{eq:kinematic_compatibility}, \eqref{eq:jump_conditions_linear_momentum}, and \eqref{eq:jump_conditions_energy} above occur when the string is amorphously quiescent on one side of the shock. In this context, for definiteness, we shall refer to such a shock as \emph{external}, while the shock described so far will be referred to as \emph{internal}.
An external shock is an attempt at formalizing the notion of \emph{continually} imparted impacts introduced in the work of Cayley~\cite{cayley:class}; as such, it is more than just an internal shock with vanishing velocity on one side. At an external shock, mass is not conserved, as the string is there in contact with a reservoir, where a shapeless deposit of mass serves as a supply of links abruptly injected one-by-one into the moving string. More generally, the moving system receives from the external reservoir supplies of mass, linear momentum, and energy, which enter the corresponding balance laws. Letting $\vel^-\equiv\zero$ and $\acc^-\equiv\zero$, and dropping everywhere the superscript $^+$ to avoid clutter, by the same reasoning leading us to \eqref{eq:kinematic_compatibility}, \eqref{eq:jump_conditions_linear_momentum}, and \eqref{eq:jump_conditions_energy}, we obtain that
\begin{subequations}\label{eq:jump_external_plus}
\begin{equation}\label{eq:jump_external_plus_velocity}
\vel+\sspeed\ut=\zero,
\end{equation}
\begin{equation}\label{eq:jump_external_plus_acceleration}
\acc+2\sspeed\vel'+\sspeed^2c\n+\ddot{s}_0\ut=\zero,
\end{equation}
\begin{equation}\label{eq:jump_external_plus_momentum}
\taut+\lambda\sspeed\vel+\Force^\ast=\zero,
\end{equation}
\begin{equation}\label{eq:jump_external_plus_energy}
\taut\cdot\vel+\frac12\lambda\sspeed v^2+W^\ast=0,
\end{equation}
\end{subequations}
where $\Force^\ast$ and $W^\ast$ denote the appropriate supplies.\footnote{In particular, equation \eqref{eq:jump_external_plus_velocity} is nothing but the statement that links are injected along the tangent to the present shape of the string. This is a necessary boundary condition for the existence of a steady solution of the dynamics of the string that preserves its shape.} Similar expressions, apart from changing $\Force^\ast$ and $W^\ast$ into their opposite, are obtained if $\vel^+\equiv\zero$ and $\acc^+\equiv\zero$. Combining \eqref{eq:jump_external_plus_velocity}, \eqref{eq:jump_external_plus_momentum}, and \eqref{eq:jump_external_plus_energy}, we see that
\begin{equation}\label{eq:jump_external_plus_consequences}
\begin{split}
\vel=v\ut,&\qquad\sspeed=-v,\\
\Force^\ast=-(\tau-\lambda v^2)\ut,&\qquad W^\ast=-\left(\tau-\frac12\lambda v^2\right)v.
\end{split}
\end{equation}
Similarly, if at a singular point the string comes instantaneously to a halt, instead of being instantaneously set in motion, equations \eqref{eq:jump_external_plus_consequences} are replaced by
\begin{equation}\label{eq:jump_external_minus_consequences}
\begin{split}
\vel=v\ut,&\qquad\sspeed=-v,\\
\Force^\ast=(\tau-\lambda v^2)\ut,&\qquad W^\ast=\left(\tau-\frac12\lambda v^2\right)v.
\end{split}
\end{equation}

Equations \eqref{eq:jump_external_plus_consequences} and \eqref{eq:jump_external_minus_consequences}, in particular, allow us to interpret $\Force^\ast$ as the \emph{continuous-impact} force envisaged by Cayley \cite{cayley:class} to describe mechanical systems in which particles of infinitesimal mass are continuously taken into ``connexion'' or are continuously lost. In both cases, an external shock is propagating backwards relative to the string at the same scalar velocity as the material in the string, so that the shock results steady in space. Thus, with the aid of \eqref{eq:jump_external_plus_consequences} and \eqref{eq:jump_external_minus_consequences}, we can also phrase in terms of external shocks the dynamics of systems with variable mass, for which Cayley~\cite{cayley:class} had proposed an \emph{ad hoc} variational principle. The complementary expressions for $W^\ast$ give the energy lost (or gained) by the string in being either set in motion or brought to a halt instantaneously. (We shall return to this in Sec.~\ref{sec:energy_balance}.)

\subsection{Shock Dissipation}\label{sec:shock_dissipation}
When the shock is internal, that is, the singular point is both followed and preceded by mass in motion, the shock dissipation $\Ws$ should depend only on the impact mechanism responsible for the abrupt change in velocity. To posit a constitutive law for $\Ws$, we seek inspiration in the laws of impact which were already introduced in 1668 by Wallis and Wren~\cite{wallis:summary}, as recounted, for example in Whittaker's treatise \cite[pp.\,234]{whittaker:treatise}.

When in a system of mass-points all impacts happen to be characterized by the same restitution coefficient $0\leqq e\leqq1$, the kinetic energy after a single impact decreases by $(1-e)/(1+e)$ times the kinetic energy of the \emph{lost} motion, the motion that would have been composed with the motion before the impact to reproduce the motion after the impact \cite[p.\,235]{whittaker:treatise}. By applying this law to the elementary transfer of mass through the shock suffered by a string, interpreted as an internal impact, we justify setting
\begin{equation}\label{eq:W_s_definition}
\Ws:=-\frac12f\lambda |\sspeed|\jump{\vel}^2,
\end{equation}
where $0\leqq f\leqq1$ will be treated as a phenomenological parameter.\footnote{Letting for a moment $f:=(1-e)/(1+e)$, I note that for a \emph{plastic} impact, $e=0$ and $f=1$, whereas, for a \emph{perfectly elastic} impact, $e=1$ and $f=0$. However, in the absence of a microscopic mechanism illuminating the origin of $\Ws$, these correspondences are purely formal and $f$ remains a constitutive parameter of the string.} In the ideal limit where $f\to0^+$, the shock is not dissipative. On the other hand, for $f=1$, the shock is maximally dissipative. In practice, for a chain $f$ should depend on both the material and shape of the constitutive links.\footnote{It might be objected that $\Ws$ could equally well be regarded as an external energy sink, instead of the energy lost in an internal impact. The latter interpretation, however, provides a better justification for the explicit law posited in \eqref{eq:W_s_definition}. Moreover, linking $f$ to an internal dissipation process makes it constitutive of the string alone, a property that seems to have some experimental ground (see Sec.~\ref{sec:discussion}).}

Equation \eqref{eq:W_s_definition} is not completely unprecedented: a special form of it can be found, for example, in (7.7) of \cite{reilly:treatment}, though there $\vel^+$ and $\vel^-$ were along one and the same direction and so the geometric ingredient introduced in \eqref{eq:W_s_definition} by $\jump{\vel}^2$ was missing. I shall return in Sec.~\ref{sec:discussion} to a possible way of determining $f$ experimentally, as I think that recently it has already been found, though perhaps inadvertently.

\subsection{Inverted Catenary}\label{sec:catenary}
\begin{figure}[]
  \includegraphics[width=.4\linewidth]{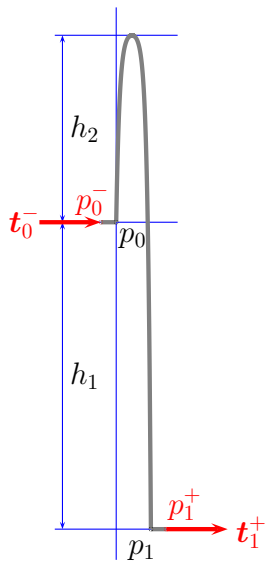}
  \caption{
  Sketch of a chain fountain. Points $p_0$ and $p_1$ designate steady internal shocks, whereas $\pzero$ and $\pone$, which are the points where the chain comes in contact with the supplying coil and the accumulating heap, respectively, designate external shocks. The drop and rise heights of the fountain relative to the supporting plane are $h_1$ and $h_2$, respectively.}
\label{fig:sketch}
\end{figure}

Before solving the balance equations for a fountain chain, we need to specify, albeit in an idealized fashion, both the pickup and putdown mechanisms that we envisage. Figure~\ref{fig:sketch} illustrates the mechanisms considered here. The points $p_0$ and $p_1$, where the chain abandons the supporting plane and where it reaches the floor, respectively, are thought of as steady internal shocks, whose kinematic compatibility with the dynamic solution is still to be established.
At points $\pzero$ and $\pone$, lying on the supporting plane and the floor, the chain comes in contact with the supplying coil and the accumulating heap, respectively, both at rest. We deliberately ignore the details about the way the chain is either coiled or heaped up: we shall be contented with assuming that at the points $\pzero$ and $\pone$ the velocity of the links constituting the chain is abruptly raised from nought or abruptly depressed to nought, respectively, so that in keeping with the notation of \eqref{eq:jump_external_plus_consequences} and \eqref{eq:jump_external_minus_consequences}, $\vel^-\equiv\zero$ at $\pzero$ and $\vel^+\equiv\zero$ at $\pone$. Thus, also $\pzero$ and $\pone$ are singular points and they can be qualified as \emph{external} shocks.\footnote{Also in accord with the attitude we take of ignoring the details about the external portions of chain they are connected with.}

The dynamics of the smooth arc of a chain fountain is governed by equation \eqref{eq:smooth_momentum_balance}, while equations \eqref{eq:kinematic_compatibility} through \eqref{eq:jump_external_minus_consequences} are to be enforced at the singular points identified above. We shall seek the solution to the problem within a special class, that of steady motions. To this end, we assume that the trajectory followed by the chain's links is invariable in time and that the spatial velocity field $\vel$ on it takes the form $\vel=v\ut$, with $v$ constant.

Projecting both sides of equation \eqref{eq:smooth_momentum_balance} along the tangent $\ut$, the principal normal $\n$ and the binormal $\bm{b}:=\ut\times\n$ to the chain's steady shape, we arrive at
\begin{equation}\label{eq:smooth_momentum_balance_components}
\tau'+f_t=0,\qquad(\lambda v^2-\tau)c=f_n,\qquad f_b=0,
\end{equation}
where $c$ is the shape's curvature and $f_t$, $f_n$, and $f_b$ are the components of $\force$ along $\ut$, $\n$, and $\bm{b}$, respectively (see also \cite{airy:mechanical}).

\begin{figure}[ht]
  \centering
  \subfigure[]{\includegraphics[width=.25\linewidth]{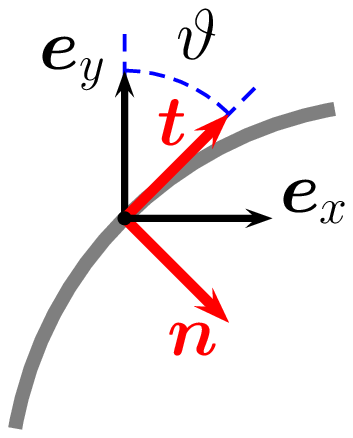}}\\
  \subfigure[]{\includegraphics[width=.46\linewidth]{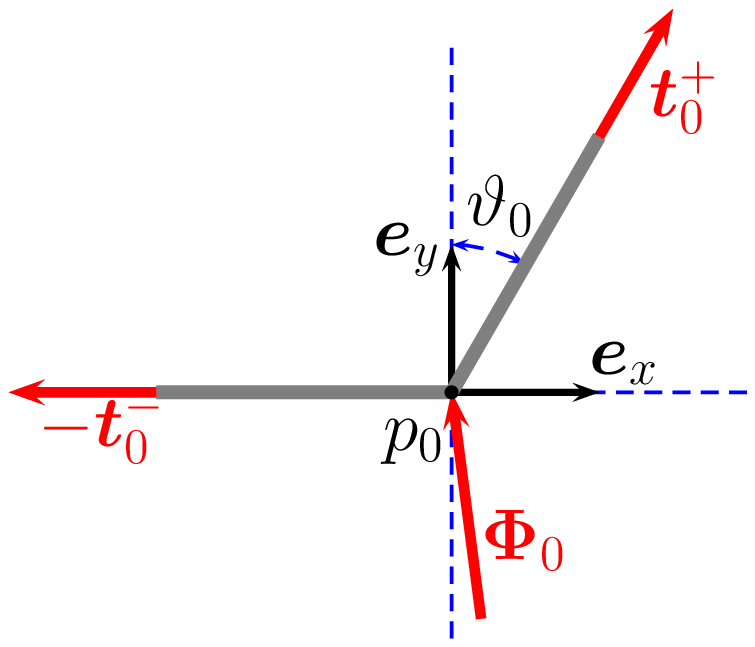}}
  \hspace{.05\linewidth}
  \subfigure[]{\includegraphics[width=.46\linewidth]{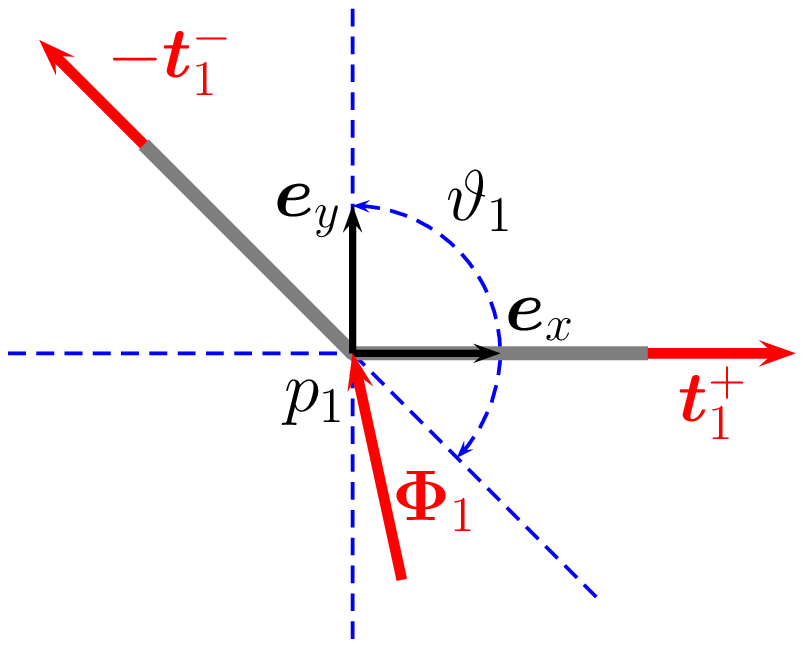}}
  \caption{
  Blowups of different significant portions of the steady shape of a chain fountain:
  (a) Generic arc with the local, movable frame $(\ut,\n)$ and a fixed, Cartesian frame $(\ex,\ey)$; (b) Arc around the pickup point $p_0$;  (c) Arc around the putdown point $p_1$. The different unit tangent vectors are described analytically by \eqref{eq:unit_tangents}. Forces $\Force_0$ and $\Force_1$ are realizations of the momentum supply  $\Force$ featuring in \eqref{eq:jump_conditions_linear_momentum}.}
\label{fig:arc_and_jumps}
\end{figure}

Letting $\force$ lie in the $(x,y)$ plane, $f_b$ vanishes identically as long as the chain's shape lies in that plane as well. Figure~\ref{fig:arc_and_jumps}(a) describes a generic arc of the chain's shape.
Denoting by $\vt$ the angle that $\ut$ makes with $\ey$, we can represent $\ut$ and $\n$ as
\begin{equation}\label{eq:movable_frame}
\ut=\sin\vt\,\ex+\cos\vt\,\ey,\quad\n=\cos\vt\,\ex-\sin\vt\,\ey,
\end{equation}
whence it follows that $c=\vt'$. Thus, as long as $c$ does not vanish, $\vt$ and $s$ can equally be employed to parameterize the chain's shape: in the setting described by Figs.~\ref{fig:sketch}, \ref{fig:arc_and_jumps}(b), and \ref{fig:arc_and_jumps}(c), $\tzero\leqq\vt\leqq\tone$. Expressing both $f_t$ and $f_n$ as functions of $\vt$, for $f_n\neq0$, we readily obtain from \eqref{eq:smooth_momentum_balance_components} that
\begin{equation}\label{eq:pre_shape_solution}
\ln|\lambda v^2-\tau|=\int\frac{f_t}{f_n}d\vt,\qquad c=\frac{f_n}{\lambda v^2-\tau}.
\end{equation}

If $\force=-\lambda g\ey$, where $g$ is the acceleration of gravity, then $f_t=-\lambda g\cos\vt$, $f_n=-\lambda g\sin\vt$, and \eqref{eq:pre_shape_solution} yields
\begin{equation}\label{eq:shape_solution}
\tau=\lambda v^2-\frac{a^2}{\sin\vt},\qquad c=\frac{\lambda g}{a^2}\sin^2\vt,
\end{equation}
where $a^2$ is a yet unknown, positive integration constant. As already remarked in \cite[p.\,64]{walton:solutions}, the shape described by \eqref{eq:shape_solution} is an \emph{inverted} catenary. Moreover, for $\tau$ not to be negative somewhere, it suffices that $\tau_1:=\tau(\tone)\geqq0$, that is,
\begin{equation}\label{eq:a^2_inequality}
a^2\leqq\lambda v^2\sin\tone.
\end{equation}
By integrating in $\vt$, with the aid of \eqref{eq:shape_solution}, the equations
\begin{equation}\label{eq:shape_presolution_x_y}
\frac{dx}{d\vt}=\frac{\sin\vt}{c},\qquad\frac{dy}{d\vt}=\frac{\cos\vt}{c},
\end{equation}
which follow form \eqref{eq:movable_frame}, we arrive at
\begin{subequations}\label{eq:shape_solution_x_y}
\begin{equation}\label{eq:shape_solution_x}
x(\vt)=\frac{a^2}{\lambda g}\left(\ln\frac{1-\cos\vt}{\sin\vt}-\ln\frac{1-\cos\tzero}{\sin\tzero}\right),
\end{equation}
\begin{equation}\label{eq:shape_solution_y}
y(\vt)=\frac{a^2}{\lambda g}\left(\frac{1}{\sin\tzero}-\frac{1}{\sin\vt}\right),
\end{equation}
\end{subequations}
which parameterize the chain's steady shape in the Cartesian plane $(x,y)$ with origin at $p_0$. Likewise, the correspondence between $\vt$ and $s$ is expressed explicitly by
\begin{equation}\label{eq:shape_soltion_s_of_theta}
s(\vt)=\frac{a^2}{\lambda g}\left(\cot\tzero-\cot\vt\right).
\end{equation}

So far we have considered both the impressed scalar velocity $v\geqq0$ and the pickup angle $0\leqq\tzero\leqq\frac\pi2$ as parameters of the solution we seek. The solution of the balance equation for linear momentum along the smooth arc of a chain fountain has identified two further parameters, $a^2$ and $\tone$, subject to the bound \eqref{eq:a^2_inequality}. In the next section, by use of appropriate boundary conditions, we shall resolve the shocks and devise a strategy to determine all four unknowns encountered here (plus two more we shall encounter there).

\subsection{Shocks Resolution}\label{sec:shock_resolution}
Combining \eqref{eq:kinematic_compatibility_velocity} with the assumption that at a shock $\vel^+=v^+\tp$ and $\vel^-=v^-\tm$, we immediately conclude that $v^+=v^-=v$ and $\sspeed=-v$. Making use of this in \eqref{eq:kinematic_compatibility_acceleration} changes the latter into an identity. As shown in Figs.~\ref{fig:arc_and_jumps}(b) and \ref{fig:arc_and_jumps}(c), $\Force_0$ and $\Force_1$ are the momentum supplies acting at $p_0$ and $p_1$, respectively, where equation \eqref{eq:jump_conditions_linear_momentum} can now be enforced in the form
\begin{equation}\label{eq:jump_conditions_linear_momentum_simplified}
\jump{(\tau-\lvtwo)\ut}+\Force=\zero,
\end{equation}
where $\Force$ is either $\Force_0$ or $\Force_1$, depending on the shock being considered. In a similar way, with the aid of \eqref{eq:W_s_definition}, at both $p_0$ and $p_1$ \eqref{eq:jump_conditions_energy} becomes
\begin{equation}\label{eq:jump_conditions_energy_simplified}
\jump{\tau}=\frac12 f\lvtwo\jump{\ut}^2.
\end{equation}

While equation \eqref{eq:jump_conditions_linear_momentum_simplified} written for both shocks determines the momentum supplies $\Force_0$ and $\Force_1$, correspondingly the jump condition \eqref{eq:jump_conditions_energy_simplified} ties $a^2$ to $\tzero$ and $\tone$ via \eqref{eq:shape_solution}. Overall, there are six unknowns that need to be determined to identify completely the steady solution we seek here, namely, $\tzero$, $\tone$, $v$, $a^2$, $\taup$, and $\taum$, where the latter two designate the tensions at the points $\pone$ and $\pzero$, respectively (see Fig.~\ref{fig:sketch}). The jump condition \eqref{eq:jump_conditions_energy_simplified} written for the two internal shocks provides only two equations: four others are missing.

As partly anticipated in Sec.~\ref{sec:catenary}, my strategy will start by treating $\tzero$ and $v$ as parameters that label all possible solutions; only later, extra conditions are to be identified which may fix them. This has the dual advantage of showing both the richness of solutions of the problem and the possibility of selecting different solutions corresponding to different conditions. Since, as will be clearer shortly below, the problem is highly non-linear, the existence of solutions in different classes will be subject to different compatibility conditions. I have privileged one class of solutions upon possible others, as I believe that that represents more closely the physical nature of the anomalous phenomenon being described by this idealized mathematical model, but the strategy proposed here could easily be applied to find solutions in other classes as well.

Thus, treating both $\tzero$ and $v$ as parameters, only two equations would be missing for the moment. One comes from the geometric condition
\begin{equation}\label{eq:geometric_condition}
y(\tone)=-h_1,
\end{equation}
which prescribes the total downfall of the chain (see Fig.~\ref{fig:sketch}). The other is the boundary condition that must be required at the point $\pone$ to reflect the deposition mechanism envisaged in the model.

I think of $\pone$ as being arbitrarily close to $p_1$, distinguished from it only for being the site of the external shock  where the links in the chain come to an abrupt halt. I imagine this as a \emph{free} deposition process, for which $\Force^\ast$ in \eqref{eq:jump_external_minus_consequences} vanishes. Thus, from \eqref{eq:jump_external_minus_consequences} we obtain the condition
\begin{equation}\label{eq:boundary_condition_tau_plus}
\taup=\lvtwo,
\end{equation}
where $\taup$ is also the tension acting on $p_1$ from the right. The same assumption, however, may not be valid at the point $\pzero$ in Fig.~\ref{fig:sketch}, also taken as arbitrarily close to $p_0$, as we have no clue as to the type of connection between every single departing link in the chain and the coil left behind it. By \eqref{eq:jump_external_plus_consequences}, we can only say that the force exerted at $\pzero$ on the chain is given by
\begin{equation}\label{eq:Force_at_pzero}
\Force^\ast_-=(\lvtwo-\taum)\ut^-_0,
\end{equation}
where $\taum$ is yet to be determined. Clearly, $-\Force^\ast_-$ is the force acted upon the residual coil by the continuously departing links.

There are two compatibility conditions that a solution must meet to be acceptable: both concern the positivity of the tension $\tau$. One is \eqref{eq:a^2_inequality}, which amounts to require that $\tau_1\geqq0$, and the other is
\begin{equation}\label{eq:tension_positiveness_taum}
\taum\geqq0.
\end{equation}

To expedite the search for solutions and to retrace  more easily in them signs of universality, it is advisable to scale all lengths to $h_1$ and all velocities to $V:=\sqrt{2h_1g}$, which represents the velocity acquired by any body falling from rest down the height $h_1$. Thus, $v$ will be replaced by
\begin{equation}\label{eq:nu_definition}
\nu:=\frac{v}{V}.
\end{equation}
Moreover, to apply to the internal shocks in $p_0$ and $p_1$ the jump conditions established in Sec.~\ref{sec:shock_equations}, it is also expedient recording here, in accord with \eqref{eq:movable_frame}, the explicit expressions for the unit tangent vectors involved in the singular points depicted in Figs.~\ref{fig:arc_and_jumps}(b) and \ref{fig:arc_and_jumps}(c):
\begin{equation}\label{eq:unit_tangents}
\begin{split}
\ut_1^+&=\ut_0^-=\ex,\\
\ut_1^-&=\sin\tone\,\ex+\cos\tone\,\ey,\\
\ut_0^+&=\sin\tzero\,\ex+\cos\tzero\,\ey.
\end{split}
\end{equation}

By \eqref{eq:unit_tangents}, \eqref{eq:boundary_condition_tau_plus}, and \eqref{eq:jump_conditions_energy_simplified}, we readily obtain that
\begin{equation}\label{eq:a^2_solution}
a^2=f\lvtwo\sin\tone(1-\sin\tone),
\end{equation}
which makes \eqref{eq:a^2_inequality} automatically satisfied. Making use of \eqref{eq:a^2_solution} and \eqref{eq:shape_solution_y} in \eqref{eq:geometric_condition}, we arrive at an equation for $\tone$ that has a unique solution $\tzero\leqq\tone\leqq\pi-\tzero$, given by
\begin{equation}\label{eq:solution_tone}
\begin{split}
\tone=&\arcsin\Big\{\frac12\Big[1+\sin\tzero\\
&\qquad\qquad\quad -\sqrt{(1-\sin\tzero)^2+4\chi\sin\tzero}\Big]\Big\},
\end{split}
\end{equation}
provided that
\begin{equation}\label{eq:chi_definition_inequality}
\chi:=\frac{1}{2f\nu^2}\leqq1.
\end{equation}
It should perhaps be noted that by \eqref{eq:nu_definition} inequality \eqref{eq:chi_definition_inequality} requires the velocity at which the chain is drawn to be sufficiently larger than $V$, by an amount that increases as $f$ decreases.

When \eqref{eq:chi_definition_inequality} is obeyed and $\tone$ is given by \eqref{eq:solution_tone}, the forces $\Force_0$ and $\Force_1$ are readily delivered by \eqref{eq:jump_conditions_linear_momentum_simplified}; their Cartesian components in the $(\ex,\ey)$ frame are found to be
\begin{subequations}\label{eq:solution_Phis}
\begin{equation}\label{eq:Phi0x}
\Phi_{0x}=-f\lvtwo(1-\sin\tzero)\left[1+\frac{\sin\tone}{\sin\tzero}(1-\sin\tone)\right],
\end{equation}
\begin{equation}\label{eq:Phi0y}
\Phi_{0y}=f\lvtwo\sin\tone(1-\sin\tone)\cot\tzero,
\end{equation}
\begin{equation}\label{eq:Phi1x}
\Phi_{1x}=-f\lvtwo\sin\tone(1-\sin\tone),
\end{equation}
\begin{equation}\label{eq:Phi1y}
\Phi_{1y}=-f\lvtwo\cos\tone(1-\sin\tone),
\end{equation}
\end{subequations}
which show how both $\Phi_{0x}$ and $\Phi_{1x}$ are negative and both $\Phi_{0y}$ and $\Phi_{1y}$ are positive, for all choices of $\tzero$ and $\tone$.

Finally, $\Force^\ast_-$ in \eqref{eq:Force_at_pzero} is determined as $\Force^\ast_-=\Phi\ex$, where
\begin{equation}\label{eq:solution_Phi_minus}
\Phi:=\lvtwo-\taum=f\lvtwo\left[\frac{\sin\tone}{\sin\tzero}(1-\sin\tone)+(1-\sin\tzero)\right],
\end{equation}
where again $\tone$ is delivered by \eqref{eq:solution_tone}. In particular, \eqref{eq:solution_Phi_minus} shows that $\Phi\geqq0$ for all values of $\tzero$ and $\chi$, and so the residual coil resting on the left of $\pzero$ is subject to the force $-\Phi\ex$, which would set it into a backward motion, if not counterbalanced by friction.

\begin{figure}[ht]
  \centering
  \includegraphics[width=\linewidth]{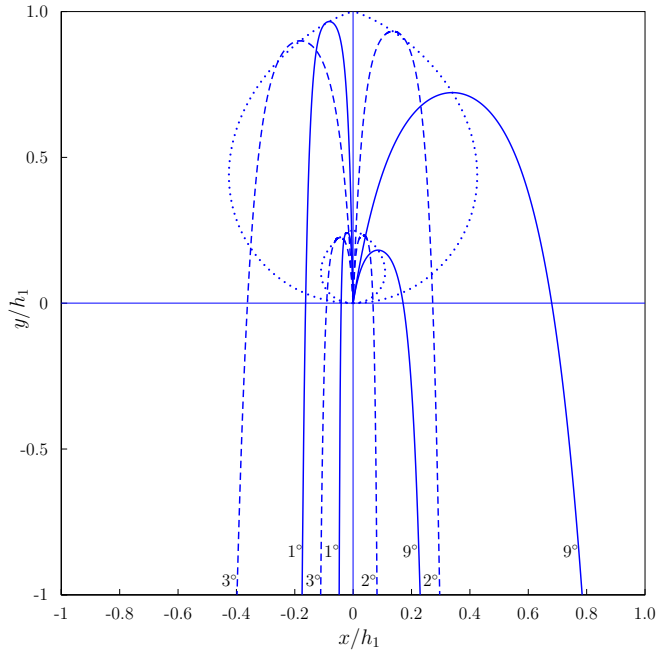}
  \caption{
  Fountain sprays according to the parametric representation \eqref{eq:solution_x_y}, drawn for $\chi=0.8$ (higher curves) and $\chi=0.5$ (lower curves). The locus of vertices of the inverted catenaries corresponding to one and the same value of $\chi$ is depicted as a dotted curve in both cases. The values of $\tzero$ are marked in degrees close to each curve. To avoid clutter, some curves have been reflected about the vertical axis, making the visual impression of a fountain even more evident.}
\label{fig:fountain}
\end{figure}

Making use of \eqref{eq:a^2_solution}, \eqref{eq:solution_tone}, and \eqref{eq:chi_definition_inequality} in the parametric representation of the inverted catenary \eqref{eq:shape_solution_x_y}, we readily arrive at
\begin{subequations}\label{eq:solution_x_y}
\begin{equation}\label{eq:solution_x}
x(\vt)=\frac{h_1}{\chi}\sin\tone(1-\sin\tone)\left(\ln\frac{1-\cos\vt}{\sin\vt} -\ln\frac{1-\cos\tzero}{\sin\tzero}\right),
\end{equation}
\begin{equation}\label{eq:solution_y}
y(\vt)=\frac{h_1}{\chi}\sin\tone(1-\sin\tone)\left(\frac{1}{\sin\tzero}-\frac{1}{\sin\vt}\right),
\end{equation}
\end{subequations}
whence, in particular, it follows that the maximum elevation $h_2$ of a chain fountain  and its width $w$ are given explicitly by (see Fig.~\ref{fig:sketch})
\begin{subequations}\label{eq:solution_height_width}
\begin{equation}\label{eq:solution_height}
h_2:=y\left(\frac\pi2\right)=\frac{h_1}{\chi}\frac{\sin\tone}{\sin\tzero}(1-\sin\tone)(1-\sin\tzero),
\end{equation}
\begin{equation}\label{eq:solution_width}
\begin{split}
w:=&x(\tone)=\frac{h_1}{\chi}\sin\tone(1-\sin\tone)\times\\ &\qquad\quad\times\left(\ln\frac{1-\cos\tone}{\sin\tone}-\ln\frac{1-\cos\tzero}{\sin\tzero}\right).
\end{split}
\end{equation}
\end{subequations}
It requires just a few computations and resort to some identities proving that \eqref{eq:solution_width} agrees perfectly with equation (17) of \cite{biggins:growth}. Figure \ref{fig:fountain} illustrates the shapes described by \eqref{eq:solution_x_y} for $\chi=0.5$ and $\chi=0.8$ and several values of $\tzero$, ranging from $1\degree$ to $9\degree$.

The solution whose shape is represented by \eqref{eq:solution_x_y} does not necessarily exist for all values of the parameters $0\leqq\tzero\leqq\frac\pi2$ and $0\leqq\chi\leqq1$, as there is no guarantee from \eqref{eq:solution_Phi_minus} that $\Phi\leqq\lvtwo$, and so \eqref{eq:tension_positiveness_taum} may be violated. We shall return to the compatibility condition imposed by this requirement shortly below, after having devised a criterion to select the kinematic parameters $(\chi,\tzero)$, which are still free.

As already remarked after \eqref{eq:solution_Phis}, the components along $\ex$ of both $\Force_0$ and $\Force_1$ oppose the chain's motion, suggesting that they are to be provided by some friction. We assume that
\begin{equation}\label{eq:Phis_friction}
\Phi_{0x}=-k_0v,\qquad\Phi_{1x}=-k_1v,
\end{equation}
where $k_0$ and $k_1$ are positive friction coefficients, possibly different from one another, and having a constitutive nature. Different physical mechanisms could be imagined to justify \eqref{eq:Phis_friction}, which however remains a criterion to select a solution out of the many parameterized above in $(\chi,\tzero)$. Other criteria could possibly be proposed, but I shall be contented with showing below that this is indeed effective.

\begin{figure}[ht]
  \centering
  \includegraphics[width=\linewidth]{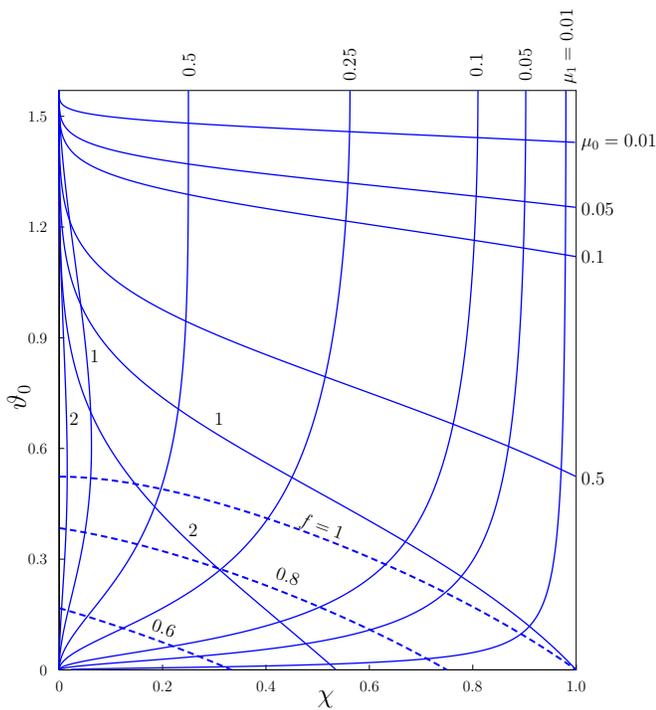}
  \caption{
  Representative curves (solid lines) of the $\mu_0$ and $\mu_1$ families described by equations \eqref{eq:selection_criterion}. The intersection between members of the two families represents an admissible solution of the chain fountain problem, provided that it falls above the appropriate curve (dashed line) in a family indexed by $f$. For $0\leqq f\leqq\frac12$, the members of this latter collapse in the origin of the $(\chi,\tzero)$ domain. For $\frac12<f\leqq1$, they grow away from the origin and cover the left corner bounded by the curve with $f=1$. The values of $\mu_0$ and $\mu_1$ are marked either outside the domain, where the corresponding curve hits the boundary, or close to the curve in the interior.}
\label{fig:Chi_Alpha}
\end{figure}

By \eqref{eq:Phi0x} and \eqref{eq:Phi1x}, the equations in \eqref{eq:Phis_friction} can be given the following dimensionless form:
\begin{subequations}\label{eq:selection_criterion}
\begin{equation}\label{eq:selection_criterion_mu_0}
(1-\sin\tzero)\left[1+\frac{\sin\tone}{\sin\tzero}(1-\sin\tone)\right]=\mu_0\sqrt{\chi},
\end{equation}
\begin{equation}\label{eq:selection_criterion_mu_1}
\sin\tone(1-\sin\tone)=\mu_1\sqrt{\chi},
\end{equation}
\end{subequations}
where
\begin{equation}\label{eq:mu_0_mu_1_definition}
\mu_0:=\frac{k_0}{\lambda\sqrt{fh_1g}},\qquad\mu_1:=\frac{k_1}{\lambda\sqrt{fh_1g}},
\end{equation}
and $\tone$ is expressed by \eqref{eq:solution_tone} in terms of $\chi$ and $\tzero$. Equations \eqref{eq:selection_criterion} are amenable to a graphical solution, aided by a bit of asymptotic analysis. Their common roots are the intersections of the curves shown in Fig.~\ref{fig:Chi_Alpha} for different values of the parameters $\mu_0$ and $\mu_1$.
As $\mu_0$ and $\mu_1$ vary, two families of curves sweep the domain $0\leqq\chi\leqq1$, $0\leqq\tzero\leqq\frac\pi2$. The members of the former run downwards and cover the whole domain as $\mu_0$ increases away from $0$: for $\mu_0\to0^+$ the curve in this family tends to the segment $\tzero=\frac\pi2$, whereas for $\mu_0\to\infty$ it tends to the orthogonal segment $\chi=0$.

The situation is a bit more intricate for the family parameterized by $\mu_1$. For $0\leqq\mu_1\leqq1$, the curves in this family run upwards in the domain $0\leqq\chi\leqq1$, $0\leqq\tzero\leqq\frac\pi2$, without covering it completely. For $\mu_1\to0^+$, the corresponding curve tends to the pair of segments $\tzero=0$ and $\chi=1$, but for $\mu_1=1$, it is tangent to the segment $\chi=0$ at both $\tzero=0$ and $\tzero=\frac\pi2$, without coinciding with it, as shown in Fig.~\ref{fig:Chi_Alpha}. Thus, for every $\mu_0\geqq0$ and $0\leqq\mu_1<1$, there is precisely one intersection between every curve in one family and every curve in the other. For $\mu_1\geqq1$, apart from the intersection at $\chi=0$ and $\tzero=\frac\pi2$, which represents a singular solution with no physical bearing, there is a meaningful intersection between representatives of the two families of curves only if
\begin{equation}\label{eq:mu_compatibility_condition}
\mu_0\geqq4(\sqrt{2\mu_1-1}-1),
\end{equation}
as can be shown through a local analysis near the point $(0,\frac\pi2)$ of the $(\chi,\tzero)$ domain.

In this domain, with the aid of \eqref{eq:solution_Phi_minus}, condition \eqref{eq:tension_positiveness_taum} can also be given a graphical representation. It can be shown that for $0\leqq f\leqq\frac12$, all points in the domain satisfy \eqref{eq:tension_positiveness_taum}, and so all intersections between $\mu_0$ and $\mu_1$ curves represent admissible solutions for the chain fountain problem. For $\frac12<f\leqq 1$, on the other hand, only the intersections that fall above a curve in a family parameterized by $f$ are admissible (see Fig.~\ref{fig:Chi_Alpha}).

In conclusion, subject to the limitations just listed, for every choice of $f$, $\mu_0$, and $\mu_1$, there is either one or no complete solution to the chain fountain problem. In Sec.~\ref{sec:discussion}, I shall compare this solution to the one already proposed by Biggins and Warner~\cite{biggins:understanding,biggins:growth} and I will comment further on the physical implications of some aspects of the solution proposed here.

\subsection{Energy Balance}\label{sec:energy_balance}
Both the internal and external shocks that feature in the chain fountain solution described above entrain an energy supply, which according to \eqref{eq:W_s_definition} is dissipative for all internal shocks, and by \eqref{eq:jump_external_plus_consequences} and \eqref{eq:jump_external_minus_consequences} may be either dissipative or productive for external shocks. The aim of this closing subsection is evaluating all the energy supplies involved in the solution and showing that, once all contributions are duly accounted for, the total energy is perfectly balanced.

We begin by computing the energy dissipated at the internal shocks steadily residing at $p_0$ and $p_1$ (Fig.~\ref{fig:sketch}), which will be denoted by $\Ws^{(0)}$ and $\Ws^{(1)}$, respectively. It follows from \eqref{eq:W_s_definition}, \eqref{eq:shape_solution}, and \eqref{eq:a^2_solution} that
\begin{equation}\label{eq:W_s_0_1}
\begin{split}
\Ws^{(0)}=-f\lvthree(1-\sin\tzero),\\
\Ws^{(1)}=-f\lvthree(1-\sin\tone),
\end{split}
\end{equation}
where, as above, $\tone$ is given as a function of $\chi$ and $\tzero$ through \eqref{eq:solution_tone}. Likewise, we apply \eqref{eq:jump_external_plus_consequences} and \eqref{eq:jump_external_minus_consequences} to the external shocks in $\pzero$ and $\pone$, where, in Cayley's language, the chain connects and disconnects, respectively, with the exterior. With the aid of \eqref{eq:boundary_condition_tau_plus} and \eqref{eq:solution_Phi_minus},we write the corresponding energy supplies as
\begin{subequations}\label{eq:W_ast_minus_plus}
\begin{equation}\label{eq:W_ast_minus}
W^\ast_-=f\lvthree\left[\frac{\sin\tone}{\sin\tzero}(1-\sin\tone)+(1-\sin\tzero)\right]-\frac12\lvthree,
\end{equation}
\begin{equation}\label{eq:W_ast_plus}
W^\ast_+=\frac12\lvthree.
\end{equation}
\end{subequations}

Finally, we compute the power $\Wa$ expended by the active external force $\force=-\lambda g\ey$. It readily follows from \eqref{eq:smooth_momentum_balance} that
\begin{equation}\label{eq:W_a}
\begin{split}
\Wa=&\int_{p_0}^{p_1}\force\cdot\vel ds=(\tau_0-\tau_1)v\\
=&f\lvthree\sin\tone(1-\sin\tone)\left(\frac{1}{\sin\tone}-\frac{1}{\sin\tzero}\right),
\end{split}
\end{equation}
where $\tau_0$ and $\tau_1$ are the internal tensions at $p_0$ and $p_1$, respectively, and use has been made of \eqref{eq:shape_solution} and \eqref{eq:a^2_solution}.

Combining together all the powers in \eqref{eq:W_s_0_1}, \eqref{eq:W_ast_minus_plus}, and \eqref{eq:W_a}, we see that they add up to zero identically, in accord with the fact  that for a steady solution the total kinetic energy $T$ is constant, and so the total balance of energy requires that
\begin{equation}\label{eq:enery_total_balance}
\dot{T}=\Wa+W^\ast_-+W^\ast_++\Ws^{(0)}+\Ws^{(1)}.
\end{equation}

A final comment is perhaps wanted about the sign of the powers obtained above. Clearly, $\Wa$ is positive, whereas both $\Ws^{(0)}$ and $\Ws^{(1)}$ are negative, as expected. On the other hand, $W^\ast_+$ is positive, indicating that energy is put into the system to produce a continuously instantaneous deposition of links; such an energy gain is exactly compensated by a loss involved in the specular pickup of links, as indicated by \eqref{eq:W_ast_minus}, where however $W^\ast_-$ may have other sign, without affecting the total balance \eqref{eq:enery_total_balance}.

\section{Discussion}\label{sec:discussion}
I had promised in the Introduction that I would relate the phenomenological coefficients $\alpha$ and $\beta$ in \eqref{eq:BW_tensions} to a single constitutive parameter. I am now in a position to fulfill that promise. Combining together \eqref{eq:shape_solution} and \eqref{eq:a^2_solution}, and comparing with \eqref{eq:BW_tensions} the expressions for the tensions $\tau_0$ and $\tau_1$ thus obtained, we readily see that
\begin{equation}\label{eq:alpha_beta_f}
\begin{split}
\alpha=&f\frac{\sin\tone}{\sin\tzero}(1-\sin\tone),\\
\beta=&1-f(1-\sin\tone),
\end{split}
\end{equation}
which express $\alpha$ and $\beta$ in terms of one constitutive parameter, $f$, and the kinematic parameters $(\chi,\tzero)$ of the solution. As shown in Sec.~\ref{sec:shock_resolution}, these latter may also be related through $\mu_0$ and $\mu_1$ in \eqref{eq:mu_0_mu_1_definition} to the friction coefficients $k_0$ and $k_1$, which unlike $f$ can be viewed as external coupling parameters. Alternatively,  it easily follows from \eqref{eq:alpha_beta_f} that
\begin{equation}\label{eq:f_theta_1}
\begin{split}
f=\frac{(1-\beta)^2}{1-\beta-\alpha\sin\tzero},\\
\sin\tone=\frac{\alpha}{1-\beta}\sin\tzero.
\end{split}
\end{equation}
By using in \eqref{eq:f_theta_1} the values $\alpha=0.12$ and $\beta=0.11$ measured in \cite{biggins:growth} together with the angle $\tzero=31\degree$, for which Fig.~5 of \cite{biggins:growth} shows the best agreement between predicted and observed shapes of the chain, we readily obtain that $f\doteq0.96$ and $\tone\doteq176\degree$. While the latter agrees qualitatively with the shapes observed in \cite{biggins:understanding,biggins:growth}, the former is a first indirect measurement of $f$.

I also alluded to the possibility that $f$ might have already been measured for some special chains. This indeed emerges from reading the work of Hamm and G\'eminard~\cite{hamm:weight} in the context of the theory presented in Sec.~\ref{sec:shocks}. Hamm and G\'eminard revisited an old problem for a special system with variable mass: the dynamics of a straight chain falling vertically under gravity and accumulating on the pan of  a scale. They observed experimentally that the free tip of the chain falls with an acceleration greater than the acceleration of gravity $g$ and proposed an equation to describe this motion featuring a dimensionless parameter, $\gamma$, which was determined by fitting the experimental data. For a ball chain, they found $\gamma\doteq0.83$ and for a loop chain, $\gamma\doteq0.95$. Moreover, they argued that ``It might be tempting to interpret $\gamma$ as characterizing the dissipation at the bottom. However, the values of $\gamma$ measured for a hard and a soft surface are the same'' and concluded that $\gamma$ ``must depend on the structure of the chain''. Now, their equation of motion (5) can also be obtained by interpreting as a dissipative shock the kink at the bottom where the chain impinges on the scale's pan. Subjecting this shock to \eqref{eq:W_s_definition}, one easily gives the balance equation of linear momentum the same form (5) of \cite{hamm:weight}, with their $\gamma$ just equal to our $f$ \cite{virga:paradoxes}.\footnote{More generally, I shall show in \cite{virga:paradoxes} how some classical chain \emph{paradoxes}, and similar ones arising in the study of systems with variable mass, can be resolved by the theory of internal dissipative shocks outlined in this paper.} That dissipation plays a crucial role in these problems is clearly witnessed by the explicit solution obtained in Sec.~\ref{sec:shocks}, in which the limit as $f\to0^+$ is highly singular and would break the required compatibility conditions, such as \eqref{eq:chi_definition_inequality}.

Another feature of the complete solution for the chain problem given here has been observed as well, though only qualitatively. It is apparent from the movies in \cite{geminard:movie} (see also \cite{biggins:growth}) that the chain being picked up exerts a backward force on the rows of chain being sucked in, a force which eventually sets the latter in motion. For such a force, now \eqref{eq:Phis_friction} delivers an explicit expression.

It is my hope that this and the other explicit details of the solution found here could prompt an experimental validation of the theory, whose main hypothesis is the existence in a chain of the internal dissipative shocks described by \eqref{eq:W_s_definition}.




\begin{thebibliography}{29}
\expandafter\ifx\csname natexlab\endcsname\relax\def\natexlab#1{#1}\fi
\expandafter\ifx\csname bibnamefont\endcsname\relax
  \def\bibnamefont#1{#1}\fi
\expandafter\ifx\csname bibfnamefont\endcsname\relax
  \def\bibfnamefont#1{#1}\fi
\expandafter\ifx\csname citenamefont\endcsname\relax
  \def\citenamefont#1{#1}\fi
\expandafter\ifx\csname url\endcsname\relax
  \def\url#1{\texttt{#1}}\fi
\expandafter\ifx\csname urlprefix\endcsname\relax\def\urlprefix{URL }\fi
\providecommand{\bibinfo}[2]{#2}
\providecommand{\eprint}[2][]{\url{#2}}

\bibitem[{\citenamefont{Mould}(2013)}]{mould:self}
\bibinfo{author}{\bibfnamefont{S.}~\bibnamefont{Mould}},
  \emph{\bibinfo{title}{Self siphoning beads}} (\bibinfo{year}{2013}),
  \bibinfo{note}{\url{http://stevemould.com/siphoning-beads/}}.

\bibitem[{\citenamefont{Hanna and King}(2011)}]{hanna:instability}
\bibinfo{author}{\bibfnamefont{J.~A.} \bibnamefont{Hanna}} \bibnamefont{and}
  \bibinfo{author}{\bibfnamefont{H.}~\bibnamefont{King}},
  \bibinfo{journal}{arXiv:1110.2360 [physics.flu-dyn]}  (\bibinfo{year}{2011}),
  \bibinfo{note}{\url{http://arxiv.org/abs/1110.2360}}.

\bibitem[{\citenamefont{Truesdell}(1960)}]{truesdell:rational}
\bibinfo{author}{\bibfnamefont{C.}~\bibnamefont{Truesdell}},
  \emph{\bibinfo{title}{The rational mechanics of flexible or elastic bodies,
  1638--1788}}, vol. \bibinfo{volume}{11/2} of \emph{\bibinfo{series}{Leonhardi
  Euleri Opera Omnia}} (\bibinfo{publisher}{Orel F\"{u}ssli},
  \bibinfo{address}{Zurich}, \bibinfo{year}{1960}), \bibinfo{note}{edited by
  Andreas Speiser, Ernst Trost, and Charles Blanc}.

\bibitem[{\citenamefont{Galieli}(1638)}]{galilei:discorsi}
\bibinfo{author}{\bibfnamefont{G.}~\bibnamefont{Galieli}},
  \emph{\bibinfo{title}{Discorsi e dimostrazioni matematiche intorno a due
  nuove scienze attinenti alla meccanica ed i movimenti locali}}
  (\bibinfo{publisher}{Elsevier}, \bibinfo{address}{Leiden},
  \bibinfo{year}{1638}), \bibinfo{note}{{O}pere (Ed. Nazionale) \textbf{8}, pp.
  39--318}.

\bibitem[{\citenamefont{Villaggio}(1997)}]{villaggio:mathematical}
\bibinfo{author}{\bibfnamefont{P.}~\bibnamefont{Villaggio}},
  \emph{\bibinfo{title}{Mathematical Models for elastic structures}}
  (\bibinfo{publisher}{Cambridge University Press},
  \bibinfo{address}{Cambridge}, \bibinfo{year}{1997}).

\bibitem[{\citenamefont{Airy}(1858)}]{airy:mechanical}
\bibinfo{author}{\bibfnamefont{G.~B.} \bibnamefont{Airy}},
  \bibinfo{journal}{Phil. Mag.} \textbf{\bibinfo{volume}{16}},
  \bibinfo{pages}{1} (\bibinfo{year}{1858}).

\bibitem[{\citenamefont{Zajac}(1957)}]{zajac:dynamics}
\bibinfo{author}{\bibfnamefont{E.~E.} \bibnamefont{Zajac}},
  \bibinfo{journal}{Bell Syst. Tech. J.} \textbf{\bibinfo{volume}{36}},
  \bibinfo{pages}{1129} (\bibinfo{year}{1957}).

\bibitem[{\citenamefont{Biggins and Warner}(2014)}]{biggins:understanding}
\bibinfo{author}{\bibfnamefont{J.~S.} \bibnamefont{Biggins}} \bibnamefont{and}
  \bibinfo{author}{\bibfnamefont{M.}~\bibnamefont{Warner}},
  \bibinfo{journal}{Proc. Roy. Soc. London A} \textbf{\bibinfo{volume}{470}}
  (\bibinfo{year}{2014}).

\bibitem[{\citenamefont{Biggins}(2014)}]{biggins:growth}
\bibinfo{author}{\bibfnamefont{J.~S.} \bibnamefont{Biggins}},
  \bibinfo{journal}{arXiv:1401.5810 [physics.class-ph]}
  (\bibinfo{year}{2014}), \bibinfo{note}{\url{http://arxiv.org/abs/1401.5810}}.

\bibitem[{\citenamefont{Hamm and G\'eminard}(2010)}]{hamm:weight}
\bibinfo{author}{\bibfnamefont{E.}~\bibnamefont{Hamm}} \bibnamefont{and}
  \bibinfo{author}{\bibfnamefont{J.-C.} \bibnamefont{G\'eminard}},
  \bibinfo{journal}{Am. J. Phys.} \textbf{\bibinfo{volume}{78}},
  \bibinfo{pages}{828} (\bibinfo{year}{2010}).

\bibitem[{\citenamefont{Grewal et~al.}(2011)\citenamefont{Grewal, Johnson, and
  Ruina}}]{grewal:chain}
\bibinfo{author}{\bibfnamefont{A.}~\bibnamefont{Grewal}},
  \bibinfo{author}{\bibfnamefont{P.}~\bibnamefont{Johnson}}, \bibnamefont{and}
  \bibinfo{author}{\bibfnamefont{A.}~\bibnamefont{Ruina}},
  \bibinfo{journal}{Am. J. Phys.} \textbf{\bibinfo{volume}{79}},
  \bibinfo{pages}{723} (\bibinfo{year}{2011}).

\bibitem[{\citenamefont{O'Reilly and Varadi}(1999)}]{reilly:treatment}
\bibinfo{author}{\bibfnamefont{O.~M.} \bibnamefont{O'Reilly}} \bibnamefont{and}
  \bibinfo{author}{\bibfnamefont{P.~C.} \bibnamefont{Varadi}},
  \bibinfo{journal}{Continuum Mech. Thermodyn.} \textbf{\bibinfo{volume}{11}},
  \bibinfo{pages}{339} (\bibinfo{year}{1999}).

\bibitem[{\citenamefont{Green and
  Naghdi}(1977{\natexlab{a}})}]{green:thermodynamics}
\bibinfo{author}{\bibfnamefont{A.~E.} \bibnamefont{Green}} \bibnamefont{and}
  \bibinfo{author}{\bibfnamefont{P.~M.} \bibnamefont{Naghdi}},
  \bibinfo{journal}{Proc. Roy. Soc. London A} \textbf{\bibinfo{volume}{357}},
  \bibinfo{pages}{253} (\bibinfo{year}{1977}{\natexlab{a}}).

\bibitem[{\citenamefont{Green and Naghdi}(1977{\natexlab{b}})}]{green:note}
\bibinfo{author}{\bibfnamefont{A.~E.} \bibnamefont{Green}} \bibnamefont{and}
  \bibinfo{author}{\bibfnamefont{P.~M.} \bibnamefont{Naghdi}},
  \bibinfo{journal}{J. Appl. Mech.} \textbf{\bibinfo{volume}{44}},
  \bibinfo{pages}{787} (\bibinfo{year}{1977}{\natexlab{b}}).

\bibitem[{\citenamefont{Green and Naghdi}(1978)}]{green:derivation}
\bibinfo{author}{\bibfnamefont{A.}~\bibnamefont{Green}} \bibnamefont{and}
  \bibinfo{author}{\bibfnamefont{P.}~\bibnamefont{Naghdi}},
  \bibinfo{journal}{J. Elast.} \textbf{\bibinfo{volume}{8}},
  \bibinfo{pages}{179} (\bibinfo{year}{1978}).

\bibitem[{\citenamefont{Green and Naghdi}(1979)}]{green:thermal}
\bibinfo{author}{\bibfnamefont{A.}~\bibnamefont{Green}} \bibnamefont{and}
  \bibinfo{author}{\bibfnamefont{P.}~\bibnamefont{Naghdi}},
  \bibinfo{journal}{Int. J. Solids Structures} \textbf{\bibinfo{volume}{15}},
  \bibinfo{pages}{829} (\bibinfo{year}{1979}).

\bibitem[{\citenamefont{O'Reilly and Varadi}(2003)}]{reilly:energetics}
\bibinfo{author}{\bibfnamefont{O.~M.} \bibnamefont{O'Reilly}} \bibnamefont{and}
  \bibinfo{author}{\bibfnamefont{P.~C.} \bibnamefont{Varadi}},
  \bibinfo{journal}{Acta Mech.} \textbf{\bibinfo{volume}{165}},
  \bibinfo{pages}{27} (\bibinfo{year}{2003}).

\bibitem[{\citenamefont{Sommerfeld}(1964)}]{sommerfeld:mechanics}
\bibinfo{author}{\bibfnamefont{A.}~\bibnamefont{Sommerfeld}},
  \emph{\bibinfo{title}{Mechanics}}, vol.~\bibinfo{volume}{1} of
  \emph{\bibinfo{series}{Lectures on Theoretical Physics}}
  (\bibinfo{publisher}{Academic Press}, \bibinfo{address}{New York},
  \bibinfo{year}{1964}).

\bibitem[{\citenamefont{M\"uller}(2007)}]{muller:history}
\bibinfo{author}{\bibfnamefont{I.}~\bibnamefont{M\"uller}},
  \emph{\bibinfo{title}{A History of Thermodynamics}}
  (\bibinfo{publisher}{Springer-Verlag}, \bibinfo{address}{Berlin},
  \bibinfo{year}{2007}).

\bibitem[{\citenamefont{Steiner and Troger}(1995)}]{steiner:equations}
\bibinfo{author}{\bibfnamefont{W.}~\bibnamefont{Steiner}} \bibnamefont{and}
  \bibinfo{author}{\bibfnamefont{H.}~\bibnamefont{Troger}},
  \bibinfo{journal}{Z. angew. Math. Phys. (ZAMP)}
  \textbf{\bibinfo{volume}{46}}, \bibinfo{pages}{960} (\bibinfo{year}{1995}).

\bibitem[{\citenamefont{Wong and Yasui}(2006)}]{wong:falling}
\bibinfo{author}{\bibfnamefont{C.~W.} \bibnamefont{Wong}} \bibnamefont{and}
  \bibinfo{author}{\bibfnamefont{K.}~\bibnamefont{Yasui}},
  \bibinfo{journal}{Am. J. Phys.} \textbf{\bibinfo{volume}{74}},
  \bibinfo{pages}{490} (\bibinfo{year}{2006}).

\bibitem[{\citenamefont{Antman}(1995)}]{antman:nonlinear}
\bibinfo{author}{\bibfnamefont{S.~S.} \bibnamefont{Antman}},
  \emph{\bibinfo{title}{Nonlinear Problems of Elasticity}}, vol.
  \bibinfo{volume}{107} of \emph{\bibinfo{series}{Applied Mathematical
  Sciences}} (\bibinfo{publisher}{Springer}, \bibinfo{address}{New York},
  \bibinfo{year}{1995}).

\bibitem[{\citenamefont{Gurtin et~al.}(2010)\citenamefont{Gurtin, Fried, and
  Anand}}]{gurtin:mechanics}
\bibinfo{author}{\bibfnamefont{M.~E.} \bibnamefont{Gurtin}},
  \bibinfo{author}{\bibfnamefont{E.}~\bibnamefont{Fried}}, \bibnamefont{and}
  \bibinfo{author}{\bibfnamefont{L.}~\bibnamefont{Anand}},
  \emph{\bibinfo{title}{The Mechanics and Thermodynamics of Contiuna}}
  (\bibinfo{publisher}{Cambridge University Press},
  \bibinfo{address}{Cambridge}, \bibinfo{year}{2010}).

\bibitem[{\citenamefont{Cayley}(1856--1857)}]{cayley:class}
\bibinfo{author}{\bibfnamefont{A.}~\bibnamefont{Cayley}},
  \bibinfo{journal}{Proc. Roy. Soc. London} \textbf{\bibinfo{volume}{8}},
  \bibinfo{pages}{506} (\bibinfo{year}{1856--1857}).

\bibitem[{\citenamefont{Wallis and Wren}(1668)}]{wallis:summary}
\bibinfo{author}{\bibfnamefont{J.}~\bibnamefont{Wallis}} \bibnamefont{and}
  \bibinfo{author}{\bibfnamefont{C.}~\bibnamefont{Wren}},
  \bibinfo{journal}{Phil. Trans. Roy. Soc. London}
  \textbf{\bibinfo{volume}{3}}, \bibinfo{pages}{864} (\bibinfo{year}{1668}).

\bibitem[{\citenamefont{Whittaker}(1937)}]{whittaker:treatise}
\bibinfo{author}{\bibfnamefont{E.~T.} \bibnamefont{Whittaker}},
  \emph{\bibinfo{title}{A Treatise on the Analytical Dynamics of Particles and
  Rigid Bodies}} (\bibinfo{publisher}{Cambridge University Press},
  \bibinfo{address}{Cambridge}, \bibinfo{year}{1937}), \bibinfo{edition}{4th}
  ed., \bibinfo{note}{reissued with Forward in the Cambridge Mathematical
  Library Series 1988}.

\bibitem[{\citenamefont{Walton and Mackenzie}(2005)}]{walton:solutions}
\bibinfo{author}{\bibfnamefont{W.}~\bibnamefont{Walton}} \bibnamefont{and}
  \bibinfo{author}{\bibfnamefont{C.}~\bibnamefont{Mackenzie}},
  \emph{\bibinfo{title}{Solutions of the Problems and Riders Proposed in the
  Senate-House Examination for 1854 by the Moderators and Examiners}}
  (\bibinfo{publisher}{Adamant Media Corporation}, \bibinfo{address}{Boston},
  \bibinfo{year}{2005}), \bibinfo{note}{replica of 1854 edition by Macmillan
  and Co., Cambridge}.

\bibitem[{\citenamefont{Virga}(2014)}]{virga:paradoxes}
\bibinfo{author}{\bibfnamefont{E.~G.} \bibnamefont{Virga}}
  (\bibinfo{year}{2014}), \bibinfo{note}{unpublished}.

\bibitem[{\citenamefont{G\'eminard}(2012)}]{geminard:movie}
\bibinfo{author}{\bibfnamefont{J.-C.} \bibnamefont{G\'eminard}},
  \emph{\bibinfo{title}{A chain falling from a table}} (\bibinfo{year}{2012}),
  \bibinfo{note}{\url{http://perso.ens-lyon.fr/jean-christophe.geminard/a_chain_falling_from_a_table.htm}}.

\end{thebibliography}

\end{document}